\documentstyle[11pt,newpasp,twoside]{article}
\begin{document}
\title{The southern dust pillars of the Carina Nebula}
\author{K. J. Brooks}
\affil{European Southern Observatory, Casilla 19001, Santiago 19, Chile,
kbrooks@eso.org}
\author{J. M. Rathborne, M. G. Burton}
\affil{School of Physics, University of New South Wales, Sydney 2052, NSW,
Australia}
\author{P. Cox}
\affil{Institut d'Astrophysique Spatiale, Universit\'{e} de Paris XI,
91405 Orsay, France}
\begin{abstract}
We present preliminary results from a detailed study towards four
previously detected bright mid-infrared sources in the southern part of the
Carina Nebula: G287.73--0.92, G287.84--0.82, G287.93--0.99 and
G288.07--0.80. All of these sources are located at the heads of giant dust
pillars that point towards the nearby massive star cluster, Trumpler 16. It
is unclear if these pillars are the prime sites for a new generation of
triggered star formation or if instead they are the only remaining parts of
the nebula where ongoing star fromation can take place.

\end{abstract}

\section{Introduction}

There is little dispute that massive stars exist in clusters and are formed
within giant molecular clouds (GMCs). All of the early sites of massive
star formation that have been identified to date are located in the
neighborhood of more evolved massive stars. In many cases, young massive
stars have been found at the heads of giant dust pillars that point toward a
more evolved massive star cluster (e.g. the elephant trunks of the Eagle
Nebula, White et al. 1999). A vigorous debate is occurring on the question
of whether this type of star formation has been triggered by the existing
cluster or is ongoing. Distinguishing between the two processes will have
implications on current massive star formation theories.  

The formation of such pillars can readily occur if a dense core within a
GMC is exposed to radiation from a nearby massive star cluster. In this
case the core would shield the column of molecular gas behind it, in a
direction pointing away from the cluster. Subsequently the more exposed
parts of the GMC would be swept up around this column or completely
irradiated away.

The Carina Nebula, at a distance of 2.2 kpc, provides an excellent
laboratory to study the environment of massive stars. The commonly held
view is that further star formation in the Carina Nebula is inhibited by
the intense feedback processes from the massive stars of Trumpler 14 and
Trumpler 16. This view has recently been challenged with a number of new
sources being identified that are good candidates for sites of current star
formation (Rathborne et al. 2001, Brooks, Storey \& Whiteoak 2001 and Smith
et al. 2000). The study by Smith et al. utilised {\it MSX} data and
revealed a region containing a number of bright mid-infrared sources
located at the tips of giant elongated emission pillars, all of which point
toward the Trumpler 16 cluster. This region is situated in the relatively
unstudied southern part of the nebula. Of the four {\it MSX} bands, the
pillars are most prominent in bands A (6.8--10.8 $\mu$m) and C (11.1--13.2
$\mu$m). Both of these bands contain several emission features arising from
polycyclic aromatic hydrocarbon molecules (PAHs) which suggests that the
pillars are bright photodissociation regions (PDRs).

\section{Results}

We have obtained $^{12}$CO(1--0) and $^{12}$CO(2--1) {\it SEST} data in the
vicinity of four of the bright mid-infrared sources detected by Smith et
al. (2000): G287.73--0.92, G287.84--0.82, G287.93--0.99 and
G288.07--0.80. Well-defined emission concentrations were detected toward
all four sources yielding relatively cool excitation temperatures of 5--20
K. For G287.93--0.99, additional observations of $^{13}$CO(2--1) were
extended over a larger area. The detected emission closely follows the
morphology of the associated mid-infrared pillar, confirming that it is a
structure of inter mixed molecular gas and dust surrounded by a PDR. A
$^{13}$CO concentration at the head of the pillar yields both an LTE and
Virial mass estimate of 700 M$_{\odot}$.

Preliminary analysis of {\it 2MASS} data in the vicinity of the
four mid-infrared sources, reveal a number of candidates for young embedded
objects. In particular, for G287.84--0.82 there is evidence for an embedded
young stellar cluster. It is tempting to conclude that this cluster is an
example of triggered star formation. Under this premise, any new stars
should preferentially form at the heads of the pillars, which are the parts
that are most exposed to the interaction of winds and UV radiation from the
existing nearby massive stars. However, one can also argue the heads of the
pillars are the remaining dense cores of the original GMC and are therefore
likely to have formed stars regardless.

With further analysis of {\it 2MASS} and {\it MSX} data we hope to
distinguish between triggered or ongoing star formation. For example, if we
find new star formation in a number of molecular concentrations throughout
the pillar then these new stars are more likely to be the result of an
ongoing star formation process which also formed the Trumpler 16 cluster.

\end{document}